\newcount\epsfscale    
\input epsf
\documentstyle[preprint,floats,aps]{revtex}
\begin{document}
\preprint{BNL\# 62759}
\title{Final Focus System for a Muon Collider: A Test Model }
\author{Juan C. Gallardo\thanks{gallardo@bnlarm.bnl.gov} and Robert B. Palmer}
\address{Center for Accelerator Physics\protect\\
Brookhaven National Laboratory\\
Upton, New York, 11793, USA}
\date{\today}
\maketitle
\begin{abstract}
The present scenario for a high luminosity 4 TeV on center of mass muon
collider requires a beta function $\beta^* \approx 3\,{\rm mm}$ at the
interaction point. We discuss a {\it test model} of a basic layout which
satisfies the requirements although it is not fully realistic. 
\end{abstract}
\section*{INTRODUCTION} The parameters for a high luminosity, high energy muon
collider are summarize  in Tb.\ref{tb1}.
The design of an extreme low-beta interaction region for a  muon
collider\cite{ref1} is non trivial and present a challenge in many ways similar
to the one encounter in the Next Linear Collider (NLC)\cite{ref2}. J.
Erwin\cite{ref3} and collaborators have designed the final focus system(FFS)
for the NLC with $\beta_x^*\approx 37\,{\rm mm},$ $\beta_y^*\approx
100\,\mu{\rm m}$ and transverse beam dimensions of $\sigma_x\approx 420\,{\rm
nm}$ and  $\sigma_y\approx 2.5\,{\rm nm}$ for the 1 TeV center of mass case.

Similarly, the latest version of the CLIC\cite{ref4} FFS also
calls for $\sigma_x\approx 90\,{\rm nm},$ $\sigma_y\approx 8\,{\rm nm}$ and
 beta functions $\beta_x^*\approx 2.2\,{\rm mm},$ $\beta_y^*\approx 0.157\,
{\rm mm}$ at the interaction point (IP) for 500 GeV in the center of mass.

Both these  designs of a FFS follow the prescription proposed by 
Brown\cite{ref5}; it consists of two telescopes with two chromatic correction
sections between them. The extremely low beta function at the IP results in the need of very strong quadrupoles which generate large
chromaticity. This chromaticity must be corrected locally and this is achieved
with two strong pairs of non-interleaved sextupoles. One pair, situated at the
position of maximum $\beta_x$  corrects the horizontal chromaticity, the other
pair at maximum $\beta_y$ corrects the vertical chromaticity. The two
sextupoles of a pair are separated by a phase advance $\phi=\pi$  $(\Delta Q
=-0.5).$ This arrangement cancels the second-order geometric aberrations of the
sextupoles thus reducing the second order tune shift by several order of 
magnitude. The bandwidth of the system is limited by the third-order 
aberrations and the
remaining second-order amplitude dependent tune shift,\cite{ref6}
\begin{eqnarray}
\Delta Q_x=& {\partial Q_x\over \partial \epsilon_x} \epsilon_x+
{\partial Q_x\over \partial \epsilon_y} \epsilon_y \nonumber \\
\Delta Q_y=& {\partial Q_y\over \partial \epsilon_x} \epsilon_x+
{\partial Q_y\over \partial \epsilon_y} \epsilon_y \label{eq1}
\end{eqnarray}
 These 
aberrations arise from: a) small phase error between the
sextupoles and the final quadruplet; b) finite length of the sextupoles. 

The residual chromaticity at the IP could be reduced by adding a number of
sextupoles at locations with nonzero dispersion, as it was suggested by
Brinkmann\cite{ref7}, which have the function of correcting locally the
chromaticity of each module. Finally, a system of octupoles could be designed
to correct the third-order aberrations. Overall, it is believed that it could be
possible to construct a system with a bandwidth of $\approx 1\,\%.$

There have been several previous attempts to design the FFS for a muon
collider\cite{ref8}, which have been summarized and compared in
ref.\cite{ref9}. 

\section {Results} 
Following the above prescription, a design
by Napoly\cite{ref10} was taken as a starting point; the final doublet used by
Napoly was replaced by a quadruplet as the final telescope\cite{ref11}. Partial
optimization of the design has been performed with the code MAD, alpha VMS
version 81.6/1\cite{ref12} Another important modification was to replace the
split sextupoles by  single elements, this simple change reduced the amplitude
dependent tune shift by an order of magnitude.

\begin{figure}[tbh]
\centering
\epsfxsize=14.0cm \epsfysize=8.0cm \epsfbox{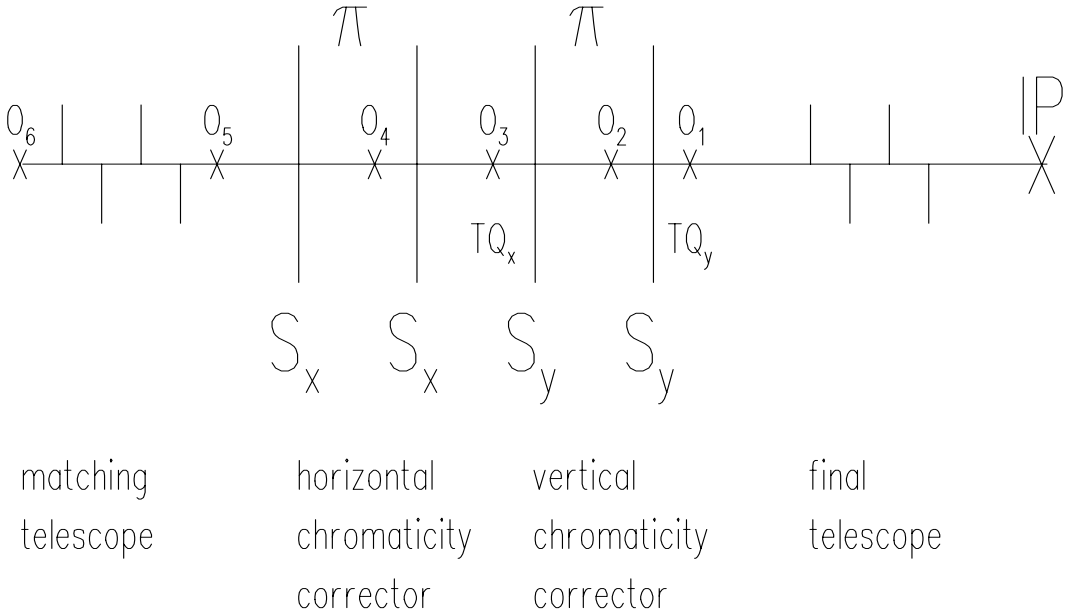}
\caption{Schematic of a FFS with extremely small beta function at the IP for a
muon collider}
\label{fg1}
\end{figure}

Starting from the interaction point (IP), there is an initial telescope with
magnification 3, ending in a focus $O_1$ (see Fig.\ref{fg1}).
 Then it follows by  FODO
cells, each with a phase advance of exactly ${\pi \over 2}.$ Intermediate foci
are generated at $O_2$, $O_3.$ Approximately midway between these foci,
vertical correction sextupoles ($S_{y1}$ and $S_{y2}$) are introduced and are
near maxima in $\beta_y.$ Then follow another similar sequence of cells with
intermediate foci at $O_4$, $O_5,$ but in these cells the sign of all
quadrupoles have been reversed. The two following sextupoles, placed between
these foci now fall on maxima of $\beta_x$ and thus serve to correct the
horizontal chromaticity $({\partial Q \over \partial \delta}).$ The horizontal
bending magnets are introduced to achieve dispersion at the sextupoles; reverse
bends are also used to reduce the dispersion between the vertical correction
sextupoles and thus avoid otherwise excessive second order dispersion $(
x\approx \delta^2).$
 The strength of the sextupoles $(S_x$ and $S_y)$ are
adjusted to minimize the first order chromaticity while trim quadrupoles 
$(TQ_x$ and $TQ_y)$ are used to minimize the second order chromaticity
$({\partial^2 Q \over \partial \delta^2}).$ The lattice is ended by a second
telescope, also with magnification 3, that could be used to match the
correction system into an arc lattice.  
The final focus system, from the matching telescope to the IP\cite{ref11}, has
a very small residual chromaticity and should be chromatically transparent when
is attached to the arc lattice.
 
The total length of the FFS is $\approx
475.8\,{\rm m}.$ The lattice consists of 44 quadrupoles, 14 sector dipoles and 
4 sextupoles. The lattice does, however, include dipoles with excessive field and
with no space between most elements.
 
\begin{figure}[tbh]
\centering
\epsfxsize=14.0cm \epsfysize=14.0cm \epsfbox{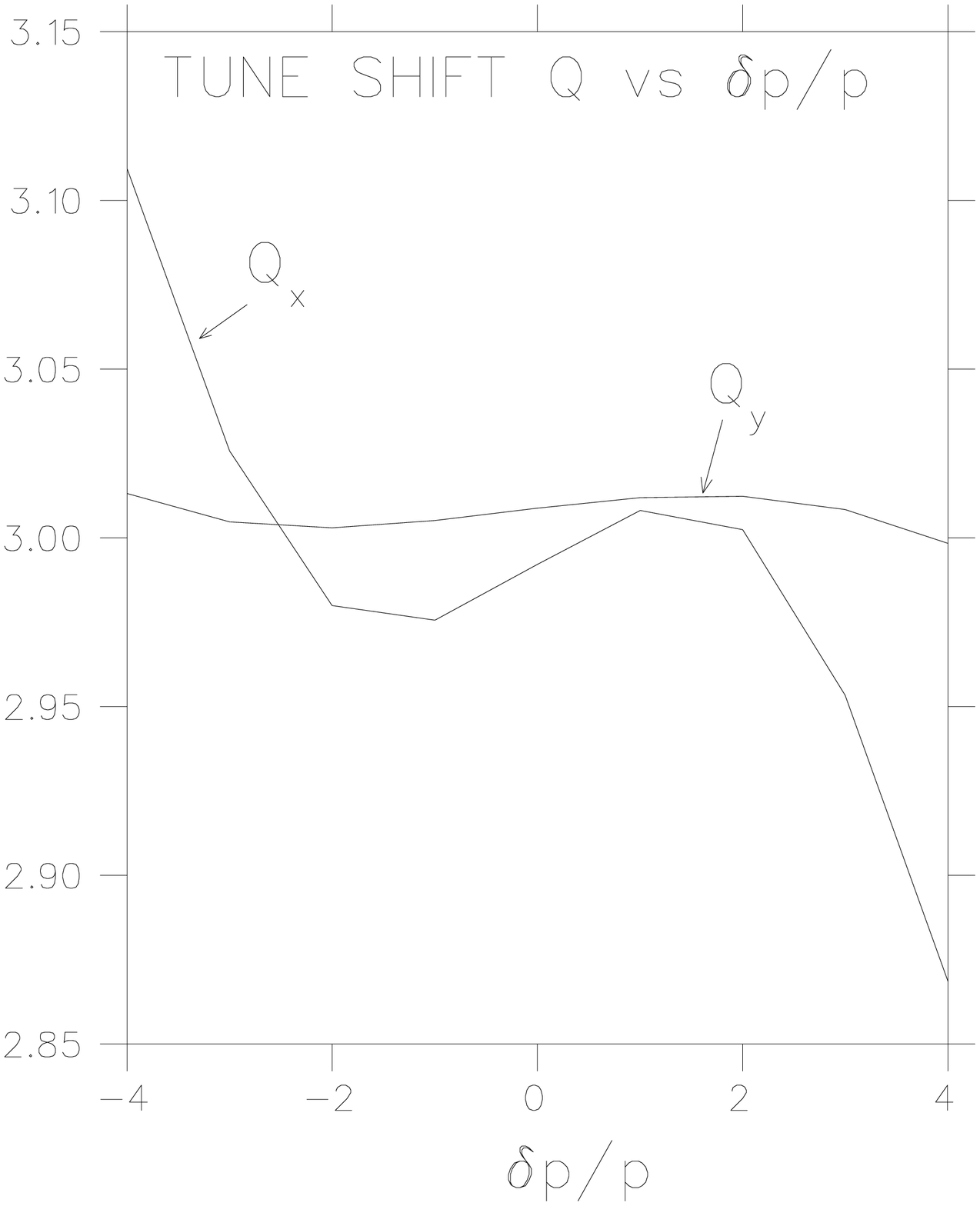}
\caption{Tune shift $Q_{x,y}$ vs  ${\delta p\over p}$ } 
\label{fg2}
\end{figure}

\begin{figure}[tbh]
\epsfxsize=14.0cm \epsfysize=14.0cm \epsfbox{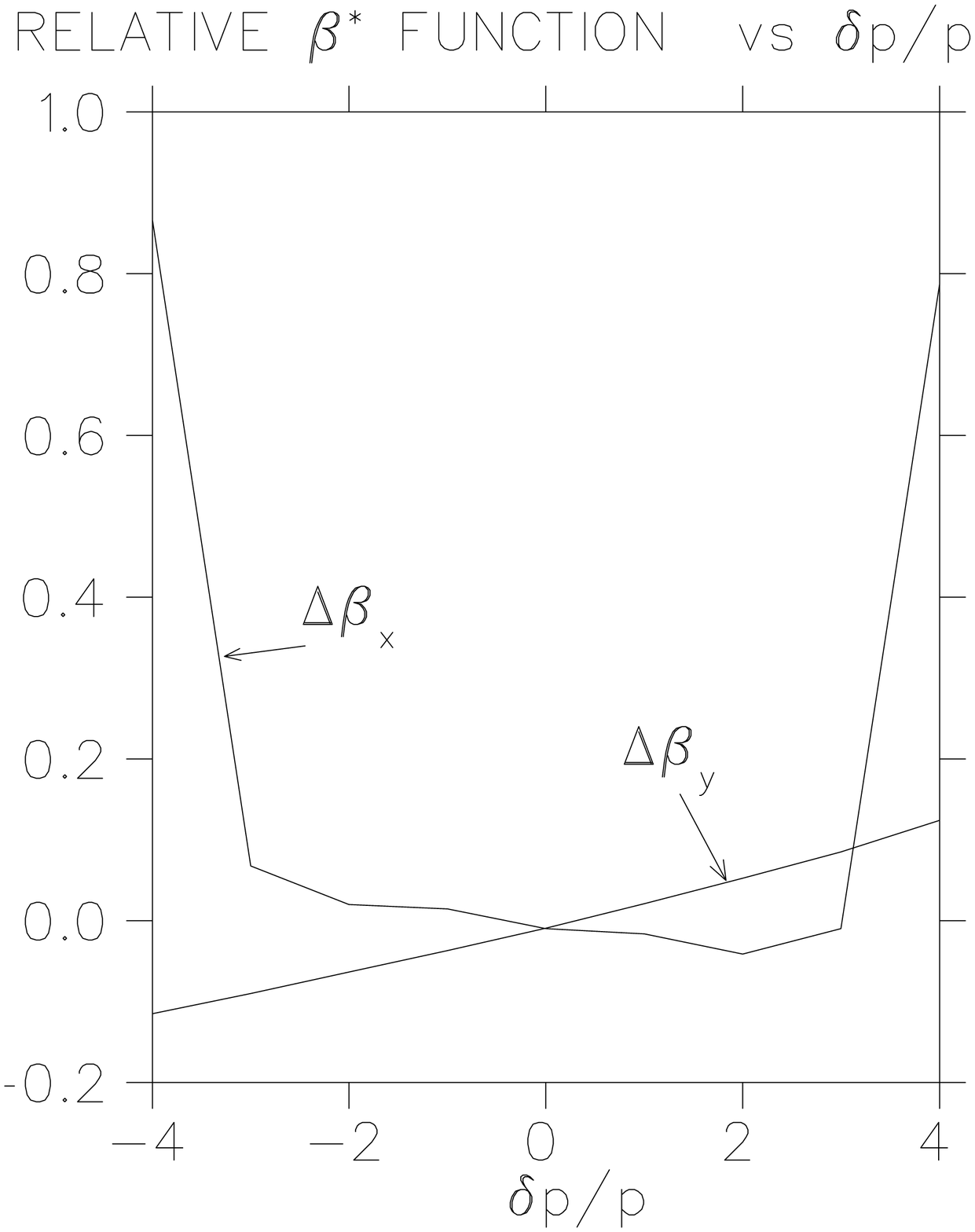}
\caption{Beta function $\beta^*$ vs ${\delta p\over p}$}
\label{fg3}
\end{figure}

\begin{figure}[tbh]
\epsfxsize=14.0cm \epsfysize=14.0cm \epsfbox{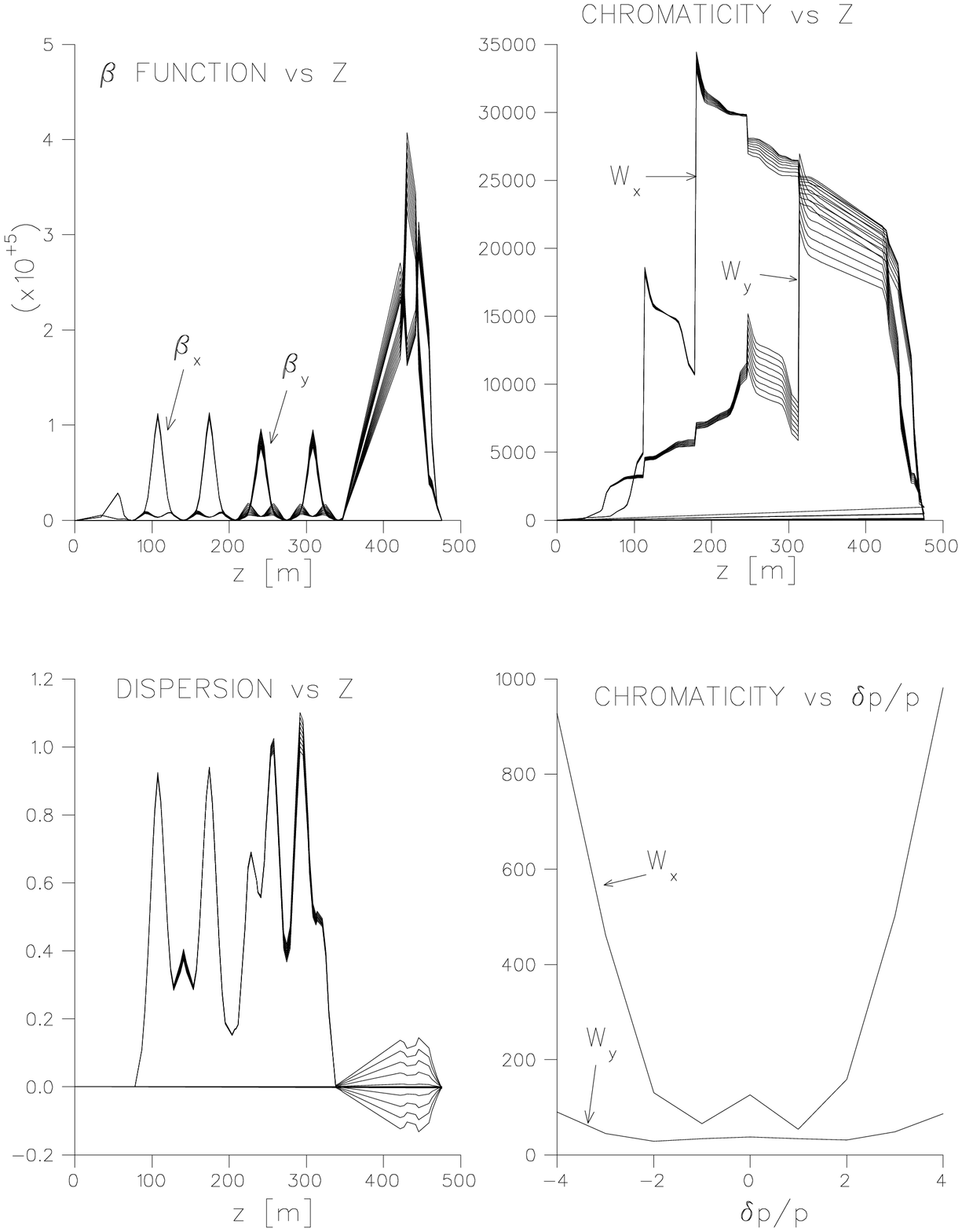}
\caption{$\beta$ functions, chromaticity and dispersion vs z for different 
momentum ($\pm 0.4\,\%$); lower right window: chromaticity vs 
${\delta p\over p}$ }
\label{fg4}
\end{figure}

The variation of the tune shift at the IP as  a function  of $\delta={\delta
p\over p}$ is shown in Fig.\ref{fg2}. 
$Q_y$ is essentially flat over a
bandwidth of $\pm 0.4\%$; $Q_x$ has obvious non-linear components, although the
variation of tune, peak to peak is less than $0.03$ within a bandwidth of $\pm
0.3\,\%.$ Likewise,  the relative $\beta^*$ variation $(\Delta
\beta^*={(\beta^*-3)\over 3})$ (Fig.\ref{fg3}) is negligible within a bandwidth
of $\pm 0.3\,\%.$ 
 The remaining figures show the beta functions as a function of the
position z (Fig.\ref{fg4}a); the chromaticity (Fig.\ref{fg4}b) and 
dispersion (Fig.\ref{fg4}c) as
function of position $z$ along the FFS and energy spread $\delta $
(Fig.\ref{fg4}d).
\section{Summary}
A {\it test model} of the FFS for a muon collider has been described. The design
satisfies the collider requirements, although it is not fully
realistic. Errors and tolerance analysis are yet to be performed as well as
tracking through the FFS to conform the achievable luminosity. 

In order to make this final focus realistic, spaces will have to be introduced
between elements and its length will have to be increased to achieve the
require dispersion without unrealistic dipole fields.

 We would like to
emphasize the need of new levels of sophistication in the correction of 
non-linear tune shift both in amplitude and momentum dependency, in storage
rings with extreme low beta functions at the IP.

\section*{ACKNOWLEDGMENTS}
This research was supported by the U.S. Department of Energy 
under Contract No. DE-ACO2-76-CH00016. (RBP) gratefully acknowledge stimulating
discussions with J. Irwin and O. Napoly; both authors thanks K.-Y. Ng and
D. Trbojevic for their helpful comments and W. Graves for his contribution in
the early stage of this work. 
\newpage

\squeezetable
\begin{table}[b]
\centering
\protect \caption{HIGH ENERGY-HIGH LUMINOSITY $\mu^+\,\mu^-$ COLLIDER}
\vspace{7mm}
\label{tb1}
\begin{tabular}{|l|l|l|}
 & 2 TeV & 250 GeV\\
\hline
Maximum c-of-m Energy [TeV]& 4&0.5   \\
Luminosity ${\cal L}$[$10^{35}$cm$^{-2}$s$^{-1}$] &1.0 & 0.05\\
Time Between Collisions  [$\mu $s]& 12 &1.5 \\
Energy Spread $\sigma_e$[units $10^{-3}$] & 2&2   \\
Pulse length $\sigma_z$[mm] & 3& 8  \\
Beam Radius at the ${\cal IP}$ [$10^{-6}$m] &2.8 & 16.0\\
Free space at the ${\cal IP}$ [m]  & 6.25 &2.0 \\
Luminosity life time [s]/No.turns & 0.02/900 & 0.003/800\\
{\it rms} Norm. emittance, $\epsilon_{x,y}^N$ [$10^{-6}$m-rad]& 50.0 &80.0\\
{\it rms} Unnorm. emittance, $\epsilon_{x,y}$ [$10^{-6}$m-rad]& 0.0026 &0.034\\
Beta Function at ${\cal IP}$, $\beta^*$ [mm]& 3 &8 \\
{\it rms} Beam size at ${\cal IP}$ [$\mu$m] &2.8 & 16.\\
Focus Pole field at IP [T] &6.0 &6.0 \\
Maximum Beta Function, $\beta_{\rm{max}}$ [km]& 400 &14 \\
Momentum Compaction, $\alpha$ [units $10^{-4}$] & 1.5 &2.0 \\
Magnet Aperture closest ${\cal IP}$ [cm] & 12 & 11.25 \\
Beam-Beam tune shift per crossing &0.04 &0.04 \\
Repetition Rate [Hz]& 15 & 15\\
RF frequency [GHz]& 3 & 1.3 \\
RF voltage [MeV]& 1500 & 140 \\
Particles per Bunch [units $10^{12}$]& 2 &4  \\
No. of Bunches of each sign & 2 & 1\\
Peak current ${\cal I}={eNc\over \sqrt{2\pi}\sigma_z}$ [kA]&12.8 & 9.6  \\
Average current ${\cal I}={eNc\over {\rm Circum}}$ [A]& 0.032& 0.25  \\
Circumference [km] & 7 &0.9 \\
Peak Magnetic Field [T] & 9 & 9 \\
Average Magnetic Field [T] & 6 & 6\\
\end{tabular} 
\end{table}


\begin{thebibliography}{99}
\bibitem{ref11}A Mad input may be obtained from the authors.
\end{thebibliography}
\end{document}